\documentclass{aa}  

\usepackage{graphicx}
\usepackage{txfonts}
\usepackage[usenames, dvipsnames]{color}

\begin{document}

\title{The RR Lyrae projected density distribution \\ from the Galactic centre to the halo}


\author{Mar\'ia Gabriela Navarro
\inst{1,2,3}
\and
Dante Minniti \inst{1,4}
\and
Roberto Capuzzo-Dolcetta \inst{2} 
\and
Javier Alonso-Garc\'ia \inst{5, 3}
\and
Rodrigo Contreras Ramos \inst{3, 6}
\and
Daniel Majaess \inst{7, 8} 
\and
Vincenzo Ripepi \inst{9}}
\institute{Departamento de Ciencias F\'isicas, Facultad de Ciencias Exactas, Universidad Andr\'es Bello, Av. Fern\'andez Concha 700, Las Condes, Santiago, Chile \\
\email{mgnavarro@uc.cl}
\and
Dipartimento di Fisica, Universit\`a degli Studi di Roma ``La Sapienza'', P.le Aldo Moro, 2, I00185 Rome, Italy 
\and
Millennium Institute of Astrophysics, Av. Vicu\~na Mackenna 4860, 782-0436, Santiago, Chile
\and
Vatican Observatory, V00120 Vatican City State, Italy
\and
Centro de Astronom\'ia (CITEVA), Universidad de Antofagasta, Av. Angamos 601, Antofagasta, Chile
\and
Instituto de Astrof\'isica, Pontificia Universidad Cat\'olica de Chile, Av. Vicu\~na Mackenna 4860, 782-0436 Mac\'ul, Santiago, Chile
\and
Mount Saint Vincent University, Halifax, Nova Scotia, Canada
\and
Saint Mary’s University, Halifax, Nova Scotia, Canada
\and
INAF-Osservatorio Astronomico di Capodimonte, Naples, Italy
}

\date{Received 22/05/2020; Accepted 08/10/2020}
\abstract{The projected density distribution of type ab RR Lyrae (RRab) stars was characterised from the innermost regions of the Milky Way to the halo, with the aim of placing constraints on the Galaxy’s evolution. 
The compiled sample ($N_{RRab}= 64,850$) stems from fundamental mode RR Lyrae variables identified by the VVV, OGLE, and Gaia surveys.  
The distribution is well fitted by three power laws over three radial intervals. In the innermost region ($R < 2.2^\circ$) the distribution follows $\Sigma_{RRab[1]}  \propto R ^{-0.94\pm 0.051}$, while in the external region the distribution adheres to $\Sigma_{RRab[2]}  \propto  R^{-1.50\pm 0.019}$ for $2.2^\circ< R <8.0^\circ$  and $\Sigma_{RRab[3]}  \propto  R ^{-2.43\pm 0.043}$ for $8.0^\circ< R <30.0^\circ$.
Conversely, the cumulative distribution of red clump (RC) giants exhibits a more concentrated distribution in the mean, but in the central $R < 2.2^\circ$ the RRab population is more peaked, whereas globular clusters (GCs) follow a density
power law ($\Sigma_{GCs}  \propto  R ^{-1.59 \pm 0.060}$ for $R<30.0^\circ$) similar to that of RRab stars, especially when considering a more metal-poor subsample ($[\rm{Fe}/\rm{H}]<-1.1$ dex). 
The main  conclusion emerging from the analysis is that the RRab distribution favours the star cluster infall and merger scenario for creating an important fraction ($>18 \%$) of the central Galactic region.  
The radii containing half of the populations (half populations radii) are $R_{H \: RRab}=6.8^\circ$ (0.99 kpc), $R_{H \: RC}=4.2^\circ$ (0.61 kpc), and $R_{H \: GCs}=11.9^\circ$  (1.75  kpc) for the RRab stars, RC giants, and GCs, respectively.
Finally, merely $\sim 1\%$ of the stars have been actually discovered in the innermost region ($R < 35$ pc) out of the expected (based on our considerations) total number of RRab therein: $N  \sim 1,562$.  
That deficit will be substantially ameliorated with future space missions like the Nancy Grace Roman Space Telescope (formerly WFIRST).} 

\maketitle

  
\section{Introduction}

The Milky Way (MW) is the only normal large spiral galaxy in the Universe for which we can study the demographics and kinematics of specific populations of stars like old and metal-poor RR Lyrae (RRL) variable stars. 
This is a limitation that reinforces the importance of the exploration of our Galaxy (see e.g. \citealt{Kunder18}).
In particular, these studies cannot be done in other local spiral galaxies like M31 and M33 because they are far enough away that crowding does not allow us to reach the innermost RRL variables. 
Other sizeable local galaxies where the RRL can be reasonably mapped all the way to their centres are the Sgr dwarf and the Large Magellanic Cloud (LMC). 
On the one hand, the LMC has a stellar halo as mapped by the RRL density distribution \citep{Drake13} and their kinematics (\citealt{Minniti03}, and \citealt{Borissova04}).
On the other hand, the Sgr dwarf galaxy is currently in an advanced state of disruption (e.g. \citealt{Majewski03}, and \citealt{Hasselquist18}), with a sizeable fraction of its RRL lost in the tails \citep{Ibata20}.
However, neither of these galaxies have been reported to have a very concentrated RRL distribution.

To study the MW formation, 
the Galactic bulge is particularly useful because the oldest populations in the MW may be found in the inner bulge regions, among the most  metal-poor stars, using tracers like old GCs \citep{beatriz16}, blue horizontal branch stars \citep{kata19}, or RRL stars (\citealt{Kunder16}, and \citealt{mm18}).

In the innermost region of the Galactic bulge we can distinguish two structures. The first is the  nuclear bulge (NB), which  can be divided into two substructures, the inner NB  with an extension of $R = 120 \pm 20$ pc \citep{Mezger}, and \citet{Launhardt}), and the outer NB with an extension of $R \sim 230$ pc. 
The second is the nuclear star cluster (NSC), which  is a very dense stellar system (\citealt{Neumayer}, and \citealt{neu20}) with  half-light radius of 2-5 pc (\citealt{rcd17}, \citealt{Geha}, and \citealt{Boker}). 
At present there are two main models to explain the formation of the NSC. 
One mechanism is the formation of the NSC stars {in situ} from a sufficient reservoir of gas trapped around the Galactic centre (see the pioneering work by \citealt{loo82}). 
The other mechanism supports the idea of the NSC formation due to the infall and merger in the central region of the hosting galaxies of massive clusters (\citealt{tre75}, \citealt{rcd93}, and \citealt{an12}).
The reliability of this mechanism has been supported by detailed $N$-body simulations that show the actual efficiency of the dynamical friction braking on the massive clusters and their subsequent merger \citep{mio08a, mio08b}.

RR Lyrae stars, in particular, are bright radial pulsators known to be good tracers of old ($>10$ Gyr) and metal-poor populations. 
RRL stars are low-mass horizontal branch stars in the core helium burning stage \citep{smith95}.
These variable stars are giant A2-F6 stars with periods between $0.2-1.1$ days, amplitudes in the optical from $0.3$ to $2$ magnitudes and in the near-IR from $0.1$ to $1$ magnitudes.
The shape of the light curve becomes more symmetrical and smaller in amplitude as we approach  the infrared wavelengths. 
Additionally, RRL stars follow a tight period--luminosity relation in the near-IR, which defines them as very precise distance indicators \citep{Catelan15}.
We can distinguish three classifications for RRL stars  (e.g. \citealt{Catelan15}): fundamental mode RRL pulsators, or RRab, which have high amplitudes and a characteristic tooth-shaped light curve;
RRc stars pulsate in the radial first overtone and have more sinusoidal light curves with lower amplitudes than the RRab stars; 
and RRd stars, which pulsate in both the fundamental mode and the first overtone, and whose light curves have a more complex shape.

The projected density distribution of RRL stars in the Galactic halo has been mapped by the classical studies of \citet{Preston59},  \citet{Saha85},  \citet{Suntzeff91},  \citet{Kinman92}, and \citet{Lee92}, among others, and more recently by different large surveys such as SDSS, PanSTARRS, and Catalina  (e.g. \citealt{Keller8}, \citealt{Sesar10},  \citealt{Sesar17},  \citealt{Carollo10},  \citealt{Akhter12}, and \citealt{Drake17}).
In the bulge--halo transition region, the RRL population was previously   mapped by \citet{Oort75},  \citet{Ibata93}, \citet{alard01}, and \cite{gran16}, among others.
The first optical microlensing surveys like MACHO \citep{Alcock93}, OGLE I \citep{Udalski93}, and EROS \citep{Aubourg93} discovered thousands of RRL stars, but the surveys were limited to a few windows of low extinction through the bulge. 
Therefore, the areal coverage was not sufficient to study the RRL projected density distribution throughout the whole bulge. 
The situation is now  much better  thanks to more modern optical surveys like OGLE IV \citep{sos14} and Gaia \citep{Gaia18dr2}, and near-IR surveys like VVV \citep{vvvminniti10},  that provide a complete bulge coverage, from the outer down to the innermost regions.
In addition, RRL stars in the Galactic centre region have  recently been discovered by \cite{dante16} and \cite{dong17} using VVV and HST data, respectively.


These surveys have different RRL star completeness, which depends on many factors that are sometimes intertwined. 
These factors range from the different instrumentations (e.g. telescope, camera, wavelength, detector) to the observational characteristics (e.g. seeing, sampling, total number of epochs, limiting magnitudes, field mapping strategy, exposure times) and analysis procedures (e.g. photometry, selection, period searching algorithms, classification), and also to external factors (e.g. stellar density, reddening,  extinction).
These tangled factors cannot be fully characterised or directly compared from survey to survey, and therefore the completeness can be widely different.
In general optical surveys are adequate for the outer bulge and halo regions, while the near-IR surveys are more suitable in the inner bulge and disk regions where the counts diminish due to heavy extinction and crowding.
For example, we have demonstrated that the microlensing events close to the Galactic centre are more efficiently discovered in the near-IR than in the optical surveys \citep{navarro17, navarro18, navarro20}.

We were motivated by the fact that the population of RRL stars is becoming more complete, thus our final aim is to study the RRL population in the entire MW, from the Galactic centre to the halo, with the best data currently available. 
RRL stars, as tracers of old populations, may help to identify the dominant mechanism  for the formation of our Galaxy, its chemical and dynamical evolution, and its inner structure.
Therefore, these variable stars play an essential role in the understanding of the structure, formation, and evolution of the MW. 
Clearly, the innermost region is the most incomplete area and the most difficult to map, and although we are interested in mapping the projected density distribution from the centre to the halo, our main interest focuses on the study of the most central region, including the formation of the Galactic bulge and NSC. 
Using RRL stars, their distribution, concentration, dynamics, and contribution to the total mass of each structure, we can understand the importance of the different proposed formation mechanisms.
Thus, the next step is to compare the observational results presented here with different models in order to explore particularly how the inner Galactic regions formed, a comparison that will be made in a follow-up paper.

Despite the obvious fact that the 3D RRL spatial distribution would give us more precise information about its population in the MW, we note the difficulties in obtaining this information from the observational and the theoretical points of view. 
The computation of the observational distribution is limited by the unknown 3D variable extinction especially in the Galactic plane and the most central region.
From the theoretical point of view, any procedure of deprojection from a 2D distribution to a 3D one is an intrinsically unreliable procedure because it implies a solution for an integral equation where the unknown 3D density distribution is convolved by a kernel function. The solution is known to be numerically unstable even in spherical symmetry, and becomes almost completely unreliable in more general non-symmetric cases.
Therefore, obtaining the 3D RRL spatial distribution is beyond the scope of this paper. Its study will need to be addressed by future datasets obtained by the next generation of telescopes.

In this article we present the observed projected density distribution of RRab stars in the inner Galaxy, within the projected  galactocentric distance $R \leq 4.8 $ kpc. 
The use of the RRab stars is motivated by the fact that, unlike RRc and RRd stars, their light curves are very asymmetrical, preventing contamination of the final sample with other variable stars such as eclipsing binaries or rotational stars. 
We examine the different RRab catalogues available, and derive for the first time parameters characterising their radial projected density distribution from the very centre to the halo, as well as the total number of RRab stars in the Galactic bulge.
We then compare the distribution with RC stars and GCs. 
RC stars are low-mass stars located in the horizontal branch of the colour-magnitude diagram (CMD), hence they also act as standard candles and distance indicators. 
The RC stars are useful for comparison with RRL stars because they are pervasive objects in the Galactic bulge and they are bright enough to be smoothly detected in this area. 
Moreover, the comparison with GCs can be used to study the Galactic formation and evolution through the analysis contribution of RRL stars that come from the disrupted GCs in the early stages of the MW. 

In Section \ref{sec:sec2} we describe the catalogues and methods used for the analysis. 
The density profile of RRab stars is presented in Section \ref{sec:sec3}. 
We discuss the presence of a bar in the RRab projected density distribution in Section \ref{sec:sec4}. 
In Section \ref{sec:sec5} we compare the distribution with RC stars and GCs. 
The bulge--halo transition is discussed in Section \ref{sec:sec6}. Section \ref{sec:sec7} discuss the role of the Nancy Grace Roman (hereafter Nancy Roman) Space Telescope in improving these results and the conclusions are presented in Section \ref{sec:sec8}.
\\ 


\section{Catalogues and method}
\label{sec:sec2}
\begin{figure}
\centering
\includegraphics[scale=0.4]{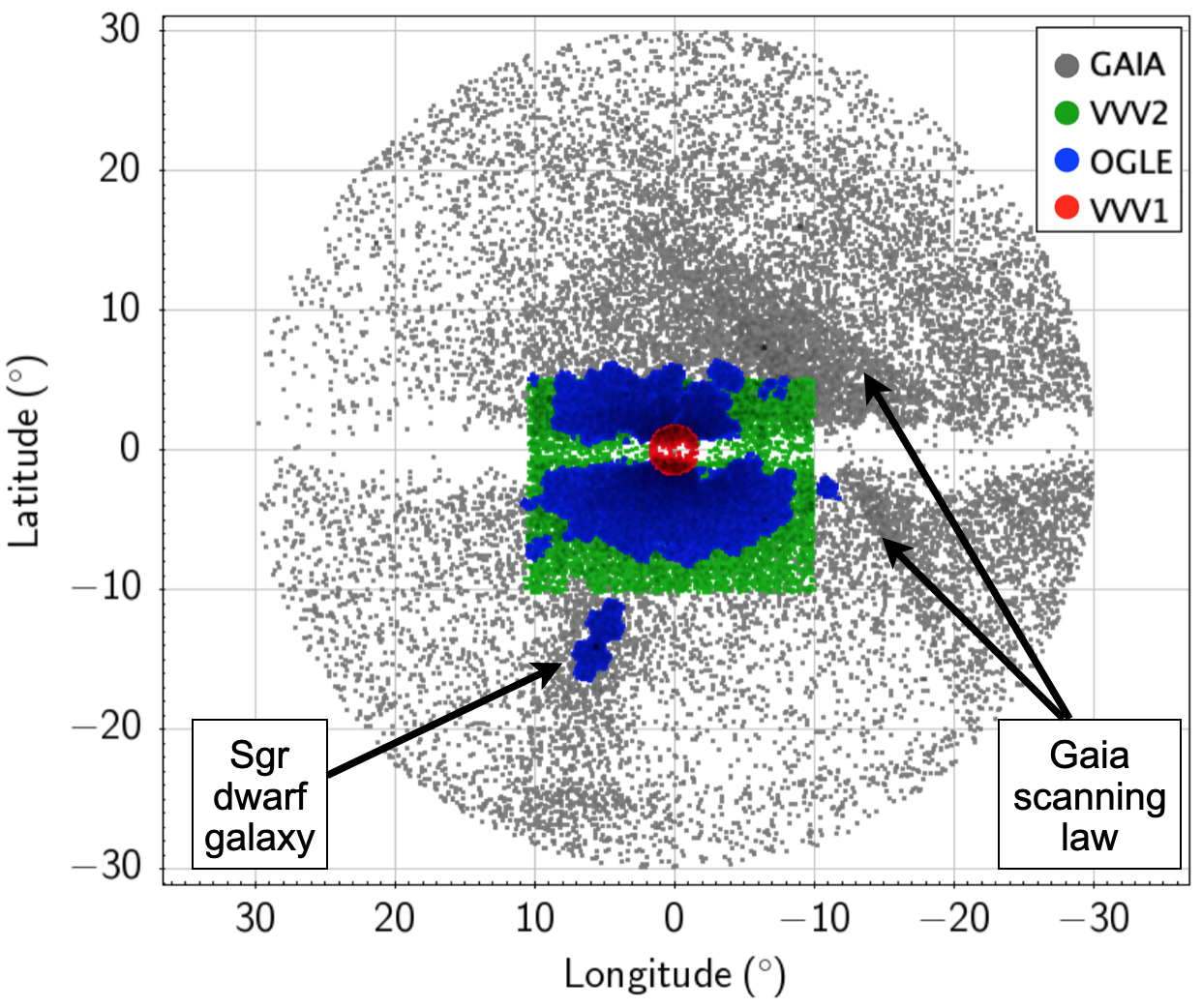}
\caption{
Map of the total sample of RRab stars used in this work. 
The red and green circles are the RRab stars discovered using the VVV1 and VVV2 catalogues, respectively; 
the blue circles are from the OGLE catalogue; and the grey circles are from Gaia DR2.
The irregular pattern for the Gaia catalogue is a consequence of its spinning and precessing pattern of observation in the sky (known as the Gaia scanning law; see e.g. \citealt{Boubert20}). 
}
\label{map}
\end{figure}

For our RRab analysis we used four catalogues from different surveys: 
i) near-IR VVV catalogues of \cite{contreras18} (VVV1); 
ii) those of \citet{dekany18},   \citet{Majaess18}, and D. Majaess private comm.;
iii) the optical catalogues of OGLE \citep{sos14}  (OGLE); 
and iv) Gaia DR2 \citep{Clementini19} (Gaia) with projected galactocentric distance $R<30.0^\circ$.
The total number of stars is $N_{RRab}=64,850$.
Table~\ref{tab:1} summarises the information of the above catalogues. 
In Fig.~\ref{map} we display the projected density distribution of the RRab population used in this analysis. 

From Fig.~\ref{map}  it is clear that the final sample is not homogeneous and is affected by observation patterns, especially in the Gaia catalogue, and structures such as the Sgr dwarf galaxy and completeness, especially in the Galactic plane and centre. 
These factors were taken into account in the following analysis to obtain the final projected density distribution.

The first step of the analysis consisted in computing the density of stars for different areas.
For this we produced a grid of equally spaced angles and variable radius using planar polar coordinates ($\rho, \theta$) , where $\rho = \sqrt{ l^2 + b^2}$ is the projected galactocentric distance ($R$) and $\theta = \tan^{-1} (b/l)$.
The grid consisted of 16 angles and a variable number in radius depending on the catalogue coverage.

Due to the strong dependence of completeness with latitude, instead of radial annuli we used squares. 
We scanned the whole area counting the stars within squares centred in every point of the grid and subsequently dividing by their area. 
The square size ($s$) increases by a factor $\Delta s$ as we move away from the Galactic centre.
The values for $s$ and $\Delta s$ depend on two variables: 
 the area covered by the catalogue, with bigger squares for catalogues covering wider areas. and
 the distance from the Galactic centre, with smaller squares in the innermost areas.
The $\rho$ increases by a factor $\Delta \rho$ proportional to the area covered by the survey. 
The main objective for this selection of the sizes is to include a significant sample of stars within the areas to obtain a reliable value for the density which varies considerably among the different catalogues and in the different areas within them.

Assuming that the VVV2 catalogue follows a similar completeness map to that obtained using VVV PSF photometry, the number of counts for VVV1 and VVV2 catalogues were corrected by incompleteness using the map of \cite{valenti16}. 
For this we interpolated the values for the completeness factor ($f_c$) obtained from \cite{valenti16} that go from 0 to 100 according to  the percentage of completeness.
We used the completeness factor of the centre of each square, so the corrected number of stars is $N_{corr}=(N_{stars} / f_c) \times 100$.

We skipped the analysis of the angles near the Galactic plane because they are more incomplete. 
For the VVV2 and OGLE catalogues the area with positive latitudes was used just to compute the density close to the Galactic plane ($b<4^\circ$), but we mostly used the area of negative longitudes, which is more complete.
For the OGLE catalogue we limited our analysis to the region with $b>-10^\circ$ to avoid the Sgr dwarf galaxy. 
For the Gaia catalogue, we used the map of duplicated sources \citep{Gaia18} to compute the density within areas less affected by the irregular sky coverage of this survey observing campaign.
Here we used areas with positive longitudes, although they are affected by the presence of the Sgr dwarf galaxy.
Therefore, during the fitting of the data the area of the distribution containing the Sgr dwarf galaxy was not considered.

Finally, the projected radial density for every catalogue was obtained as the angular average of the density at a specific $R$.
The errors for each value of $R$ corresponds to the standard deviation assuming a normal distribution, and are affected by the number of angular bins used in calculating the mean density.

There are different ways of normalising different surveys. 
Rather than normalising to one single region (e.g. Baade's window), we chose to normalise the different distributions so that they smoothly fit each other in the overlapping regions. 
In doing so, this scaling is less prone to incompleteness or systematics, and ensures better statistics.
To this end, we compared the number of counts in a set of control fields located in overlapping areas at a specific latitude and variable longitude.
The average of the set corresponds to the completeness between two catalogues.
Here we included the completeness correction for VVV1 and VVV2. 
Assuming OGLE as the most complete catalogue, the comparison with the VVV2 catalogue in the control field centred at $(l,b)=(0.0^\circ , -6.0^\circ) $ gives a $78 \%$  completeness for the latter catalogue. 
For  VVV1 we obtained a $35 \%$ completeness according to the control field located at $(l,b)=(0.0^\circ, -1.35^\circ)$. 
For the Gaia catalogue we used the control field located at $(l,b)=(0.0^\circ , -8.5^\circ)$ with a result of $35 \%$ and  $27 \%$  completeness compared with the VVV2 and OGLE catalogues, respectively.
We scaled the density profile by these factors and computed a unified density profile covering the whole range from $0.2^\circ$ to $30.0^\circ$. 

There are several reasons why the Gaia catalogue is the most incomplete one. In crowded regions of the Galactic bulge, apart from being affected by extinction and crowding, the Gaia spacecraft data rate download is limited. In the whole area the main limitation is the Gaia scanning law. There are regions with a few epochs where the period cannot be well constrained, meaning we are missing many variable stars. 
The Specific Objects Study (SOS) pipeline used to validate and characterize the Gaia RRL stars and Cepheids is described in detail in \cite{Clementini16}, and the modified version applied to DR2 is presented in \cite{Clementini19}. 
The pipeline begins reducing the sample by imposing a restriction on the number of observations for each time series in order to use the well-sampled light curves to accurately calculate the variability parameters. 
Then it applies a series of processing modules using a non-linear Fourier analysis and Fourier decomposition and relations as the period--luminosity and period--amplitude relation to select the final sample and obtain the main parameters of the variable stars.


\section{Density profile}
\label{sec:sec3}
\begin{figure}
\centering
\includegraphics[scale=0.55]{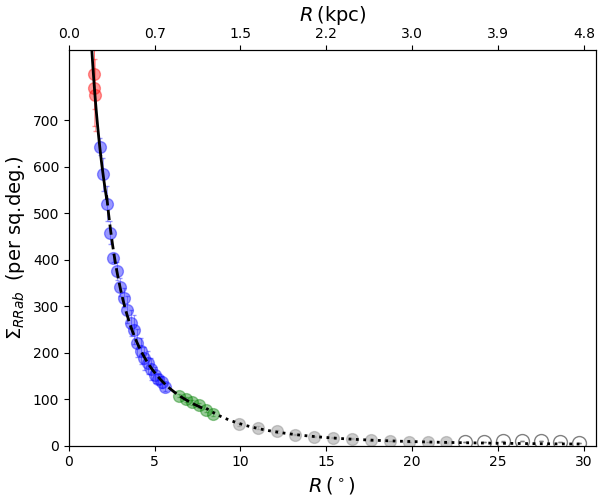}
\caption{
Projected density distribution of RRab stars.
The colour-coding is the same as in  Fig.~\ref{map}. 
The solid black line shows the power-law fit for the innermost area of $R<2.2^\circ$.
The black semi-dashed line and black dotted line show the power law for $2.2^\circ<R<8.0^\circ$  and $8.0^\circ<R<30.0^\circ$, respectively.
The open circles are the expected location of the Sgr dwarf galaxy ($R>23^\circ$), so these data points are not included in the fitting procedure.
The radius is at half maximum density $R_{\Sigma \: RRab} <1.7^\circ$ ($R_{\Sigma \: RRab}<0.25$ kpc). 
The top axis shows the distance in kiloparsecs. 
The conversion from degrees to kiloparsecs is made assuming that the distance to the Galactic centre is $R_{\odot} = 8.33$ kpc from \cite{dekany13}. 
}
\label{dd}%
\end{figure}

\begin{figure}
\centering
\includegraphics[scale=0.55]{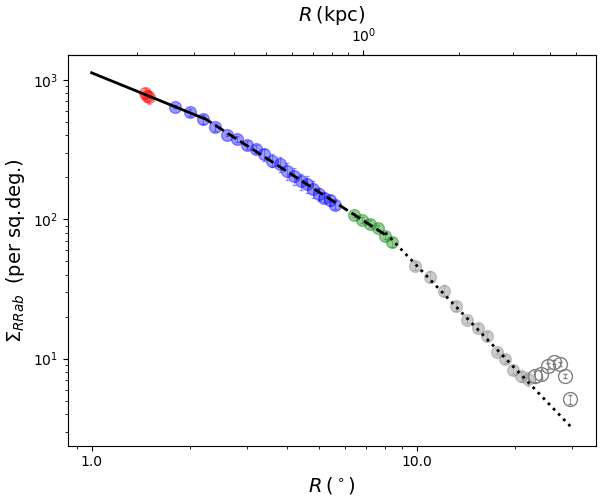}
\caption{
Projected density distribution of RRab stars in logarithmic scale together with the log-linear fits.
The colour-coding is the same as in Fig.~\ref{map}. 
The solid black line shows power-law fit for the innermost area of $R<2.2^\circ$ with a slope of $m_{RRab[1]} =-0.94 \pm 0.051$. 
The black semi-dashed line shows the best fit for $2.2^\circ<R<8.0^\circ$ with a slope $m_{RRab[2]} =-1.50\pm 0.019$. 
The best fit for $8.0^\circ<R<30.0^\circ$ corresponds to the black dotted line with slope $m_{RRab[3]} =-2.43\pm 0.043$. 
}
\label{ddlog}%
\end{figure}

The density profile varies with the projected galactocentric distance from the centre ($R \: (^\circ)$).  
As an approximation, if we want to express the result in kiloparsec we use $R \:  (\mathrm{kpc}) = R_{\odot} \tan{R} \:(^\circ)$,
where $R_{\odot} = 8.33$ kpc is the Galactic centre distance to the Sun \citep{dekany13}. 
The radial projected density distribution obtained with normal and logarithmic axes are shown in Fig.~\ref{dd} and Fig.~\ref{ddlog}. 

Due to the few data points in the innermost interval ($R<2.2^\circ$), we fitted the data to both, a linear expression and a power law.
Both the linear distribution and power law go all the way to the centre where the detection of RRL stars is not yet possible with the available telescopes. 
In this case the power law is the one that best fits the data; therefore, the linear expression obtained $\Sigma_{RRab[1]} =   (-379.6 \: R + 1336)$ per sq.deg. for $R<2.2^\circ$ is presented here for the sake of completeness, and it is used only as a lower limit to project the star population in the centre of the MW where the power law diverges.

The results in the central region are in agreement with those obtained by \cite{contreras18}. 
They estimated $\Sigma_{RRab} \sim 1,000$ per sq.deg. at $R = 1.6^\circ$; this number is similar to the $\Sigma_{RRab} \sim 719$ per sq.deg. and $\Sigma_{RRab} \sim 728$ per sq.deg. obtained using the power-law fit and linear fit, respectively.

A good fitting formula is composed of three logarithmic expressions over three radial intervals.
The best fits obtained for the different angular distances are the following;

\begin{equation}
\log (\Sigma_{RRab}) =\begin{cases}
    (-0.94 \pm 0.051)  \:  \log R + 3.0, \qquad  & \!  \! \! \!  \! \! \! \! \! \! \! \! \! \! \text{for $0.0^\circ< R <2.2^\circ$}, \\
   (-1.50 \pm 0.019)  \: \log R + 3.2, &  \!  \! \! \!  \! \! \! \! \! \! \! \! \! \!   \text{for $2.2^\circ< R <8.0^\circ $},\\
   (-2.43  \pm 0.043) \:  \log  R + 4.1, &  \!  \! \! \!  \! \! \! \! \! \! \! \! \! \!  \text{for $8.0^\circ< R <30.0^\circ$}.
  \end{cases}
\end{equation}

Previous studies show that the projected density distribution is not spherical but flattened along the Galactic longitude (\citealt{pietrukowicz15}, and \citealt{Minniti98}). 
For the sake of completeness, we computed the projected density distribution including a flattening factor ($f$). 
The method used is the same explained in Section \ref{sec:sec2}, but including a grid of flattening factors $f=R_b/R_l$ with values between $0.5 < f < 1$,  again using  planar polar coordinates ($\rho, \theta$) , where $\rho = \sqrt{ l^2 + (f b)^2}$ is the projected distance from the centre assuming a flattened distribution that correspond to the semi-major axis ($\rho = R_l$) and $\theta = \tan^{-1} (f b/l)$.
The flattening factor that minimises the sigma on the distribution has a value of $f=0.6$ in agreement with previous estimations such as the flattening factor obtained by \cite{pietrukowicz15} of $f = 0.66 \pm 0.03$ computed using OGLE data.
The projected density distribution again follows three power laws over three radial intervals: $\Sigma_{RRab[1f]}  \propto \rho ^{-0.80 \pm 0.015}$ for $\rho < 3.8^\circ$, $\Sigma_{RRab[2f]}  \propto  \rho^{-1.61\pm 0.009}$ for $3.8^\circ< \rho <11.0^\circ$  and $\Sigma_{RRab[3f]}  \propto  \rho ^{-1.98\pm 0.022}$ for $11.0^\circ< \rho <30.0^\circ$.
These distributions are similar to those listed in Eq. 1 obtained adopting a spherical distribution and so  the conclusions of this article are unchanged.

Based on Fig.~\ref{ddlog}, the RRab bulge density law smoothly merges with that of the halo, i.e.  there is no sharp bulge-halo transition in the RRab distribution, as previously studied by other authors (\citealt{Alcock1998}, and \citealt{dante99}), without a significant step or slope change in the region between $R=2 - 3$ kpc ($R=14^\circ - 20^\circ$).
The transition is visible when the analysis and comparison with RC are done in Section \ref{sec:sec6}. The distribution in the halo follows a steeper, smooth power law, as it was previously known (\citealt{Keller8},  \citealt{Drake17}, and \citealt{Akhter12}).

The expected number of RRab stars in the inner region, i.e. the number of RRab inside the projected circular shell of radius $R = 100$ pc, is $N \sim 4,613$ stars. 
For the most central region ($R <35$ pc) where \cite{dante16} and \cite{dong17} found the first five RRab stars, we estimate $N \sim 1,562$ RRab stars. 
For the Milky Way NSC, assuming a half-light radius of $R  = 4.2 \pm 0.4$ pc \citep{sch14}, we expect to find $N \sim 168$ RRab stars. 
This number is significantly greater than the $N \sim 40$ RRab stars proposed by \cite{dong17} obtained using a completely different method.

The radius at half maximum density ($R_\Sigma$) is computed using the central linear expression to avoid the divergence of the power-law fit in Eq. 1. 
Therefore, we propose the value for the radius at half maximum density $R_{\Sigma \: RRab} <1.7^\circ$ ($R_{\Sigma \: RRab}<0.25$ kpc) as an upper limit.
Nonetheless, due to the steep gradient in this area, the values of $R_{\Sigma} $ do not change considerably. 
The radius $R_{\Sigma  \: RRab}$ obtained contains $\sim 4.5 \%$ of our sample; this means that we are missing an important quantity of RRab stars in the innermost regions due to the heavy extinction and severe crowding.
Additionally, the most incomplete part of the distribution is the innermost area, so the slope can be even more pronounced. 
Our finding is that the exponential density profile continues up to $\sim 2.2^\circ$ from the Galactic centre. 
This means that the RRab population is concentrated in the inner region of the MW.

\begin{figure}
\centering
\includegraphics[scale=0.55]{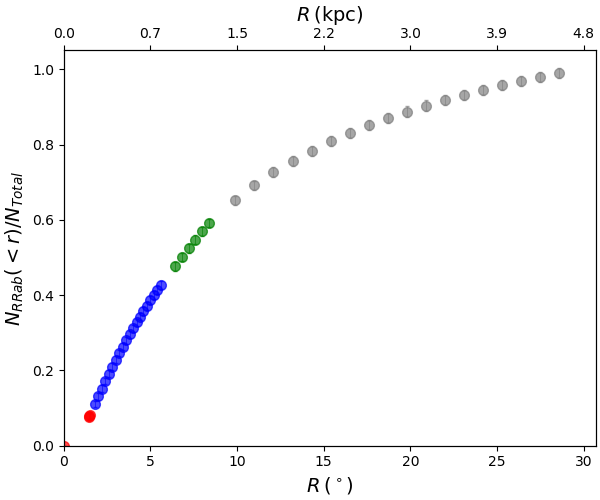}
\caption{
Normalised cumulative distribution of RRab stars.
The colour-coding is the same as in Fig.~\ref{map}. 
Fifty per cent of the sample is contained within $R_{H \: RRab}=6.8^\circ$  ($R_{H \: RRab}=0.99$ kpc).
}
\label{cd}
\end{figure}

The cumulative distribution is computed using the data-points of the density profile of RRab shown in Fig.~\ref{dd} as a factor multiplied by the area of the circle shell for the corresponding bin. 
As a double check we use the best fit for the density profile, so the cumulative distribution follows 

\begin{equation}
N(<R_f) = \int_{0}^{R_f}  \Sigma_{RRab}(R)\,2 \pi R\, dR,
\end{equation}where $N(<R_f)$ is the number of objects in the projected circular shells from the Galactic centre to $R_f$.
In both cases the distributions follow the same behaviour reaching a maximum number of $N \sim 68,000$ RRab stars within $R<30.0^\circ$.
The cumulative distribution obtained with the integral grows slightly faster than the distribution computed from the data points.
Figure~\ref{cd} shows the normalised cumulative distribution of RRab stars obtained directly from the density profile. 

From the cumulative distribution we obtain a total number of $N =67,753$ RRab stars within $R<30.0^\circ$. 
If we limit the area to the Galactic bulge ($R< 10^\circ$) we expect to find $N = 46,482$ RRab stars.
The half-population radius ($R_H$) is defined here as the radius containing half of the RRab population. 
For RRab stars the half-population radius obtained is $R_{H \: RRab}=6.8^\circ$  ($R_{H \: RRab}=0.99$ kpc).

\section{Is there a bar in the RRab distribution?}
\label{sec:sec4}

The projected density distribution of RRab is very concentrated indeed, with radius at half maximum density $R_{\Sigma \: RRab}<0.25$ kpc. 
A surprising difference arises when comparing the RRL surface distribution in the sky with the distribution along the line of sight. 
We measure $FWHM \sim 0.5$ kpc for the surface distribution.
On the other hand, \cite{pietrukowicz15} find a larger line of sight $FWHM \sim 2$ kpc. 
We can suggest two explanations for this significant difference.

The first is due to the RRL distribution being truly triaxial, with axial ratios (2.0, 1.0, 0.66) and with the axes pointing along the Galactic x, y, z axes, which seems a fortunate coincidence, though not impossible.
If this is the case, given that the bulge--halo transition is quite smooth, it would be interesting to know if the halo itself is as elongated, continuing the same shape.
The second explanation is that the line of sight distribution may be artificially inflated by the distance errors due to heavy extinction in the inner regions for example.

The real answer is probably due to a combination of the two effects, and this remains  an unsolved issue that clearly needs further study.


\section{Comparison with red clump giants and globular clusters}
\label{sec:sec5}
We have also compared the RRab projected density distribution with the distributions of the RCs obtained from the VVV near-IR photometry \citep{alonso18} and from 2MASS \citep{Cutri03}, and 
with those of the Galactic GCs from the catalogues of \cite{harris10} and \cite{Baumgardt}.

To obtain the VVV RC sample, we used the reddening-corrected Wesenheit magnitude, defined as 
\begin{equation}
W_{k_s} = K_s - 0.428 \: (J - K_s).
\end{equation}
The RC sample is obtained selecting the stars within the  area of the CMD limited by $11.5 <W_{k_s}< 13.0$ and $0.7 <(J - K_s)< 4.5$.
Then the VVV RC catalogue was corrected for incompleteness using the map of \cite{valenti16}. 

The number of RC sources for the 2MASS and VVV catalogues are $N = 4,351,918$ and $N = 19,982,084$, respectively. The catalogues of GCs contain in total $N = 101$ objects within $R<35^\circ$.

The projected density distribution and cumulative distribution were obtained following the same procedure described in detail in Section \ref{sec:sec2}.
For the RC sample of the 2MASS catalogue we computed the density profile in three angular directions: towards positive latitudes, towards  negative latitudes, and towards $\theta = 0.78$ rad ($\theta = 45^\circ$) counting from the first quadrant of Galactic coordinates $(l,b)$ where  longitude and latitude are both positive.
Therefore, the 2MASS RC catalogue consists of the stars in those directions used as representative samples of the entire area.

\begin{figure}
\centering
\includegraphics[scale=0.55]{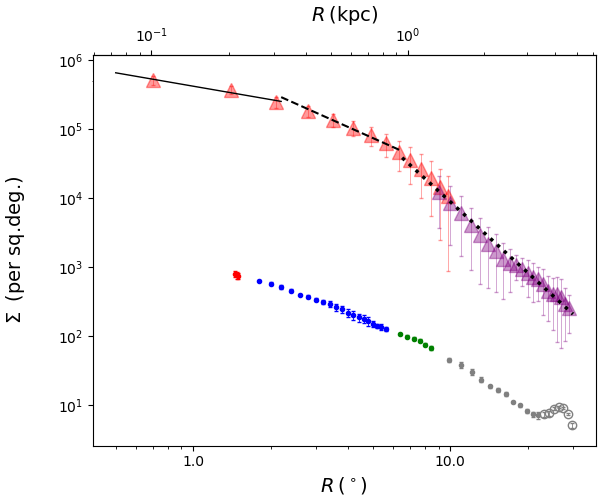}
\caption{
Projected density distribution for the RC population (triangles) and RRab (circles). 
The colours show the different catalogues used. 
The red and purple triangles correspond to the VVV and 2MASS surveys, respectively. 
The colours of the RRab distribution are the same as in Fig.~\ref{map}. 
The solid black line shows the power-law fit for the innermost area of $R<2.2^\circ$ with a slope of $m_ {RC[1]} = -0.64\pm 0.133$. 
The black semi-dashed line shows the best fit for $2.2^\circ<R<6.5^\circ$ with a slope $m_{RC[2]} =-1.66\pm 0.100$. 
The best fit for $6.5^\circ<R<30.0^\circ$ corresponds to the black dotted line with slope $m_{RC[3]}  =-3.41 \pm 0.075$. 
}
\label{rc}
\end{figure}

Figure~\ref{rc} shows  the RC and RRab  (from Fig.~\ref{ddlog}) density profiles.
The red and purple triangles correspond to the VVV and 2MASS surveys, respectively.
In the range of distances considered here, the RRab density increases towards the centre by two orders of magnitude, while the density of RCs does so by more than three orders of magnitude.

The RC projected density distribution follows a triple power law; 
in the outer region the distribution follows $\Sigma_{RC[2]}  \propto R^{-1.66 \pm 0.100}$ and $\Sigma_{RC[3]}  \propto R^{-3.41 \pm 0.075}$ for $2.2^\circ<R<6.5^\circ$ and $6.5^\circ<R<30.0^\circ$, respectively.
For the central part ($R<2.2^\circ$) the distribution follows the power law $\Sigma_{RC[1]}  \propto R^{-0.64\pm 0.133}$.

The radius at half maximum density  is $R_{\Sigma \: RC} <1.6^\circ$ ($R_{\Sigma \: RC}<0.24$ kpc). 
As for the RRab distribution, $R_{\Sigma \: RC}$ is computed using the linear fit for the density profile. 
If the density profile follows a power law in the innermost area, the density at half maximum would be even closer to the Galactic centre. 
The radius at half maximum density for the RC distribution is slightly smaller than the value obtained for the RRab distribution ($R_{\Sigma \: RRab} <1.7^\circ$); however, from the density profile we find that the RC follow a shallower power law ($\Sigma_{RC[1]}  \propto R^{-0.64\pm 0.133}$) than the RRab ($\Sigma_{RRab[1]}  \propto R^{-0.94 \pm 0.051}$) in the inner region of the Galactic bulge ($R<2.2^\circ$). 

\begin{figure}
\centering
\includegraphics[scale=0.55]{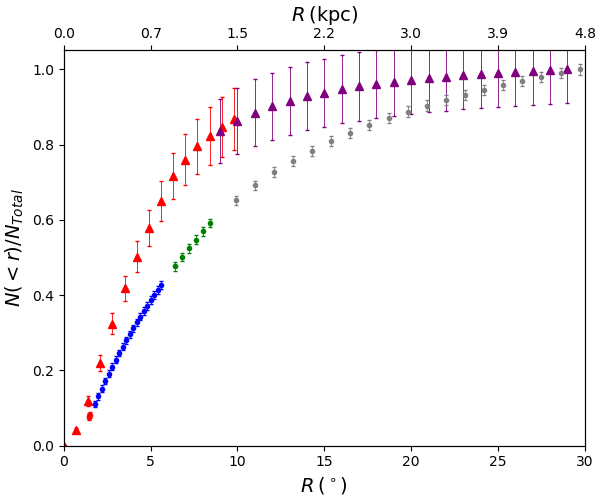}
\caption{
Cumulative distribution for the RC (triangles) and RRab (circles) populations. 
The red and purple triangles correspond to the VVV and 2MASS surveys, respectively. 
The colours of the RRab distribution are the same as in Fig.~\ref{map}. 
The half-population radius for the RC population is $R_{H \: RC}=4.2^\circ$ ($R_{H \: RC}=0.61$ kpc).
}
\label{rc_cum}
\end{figure}

Figure~\ref{rc_cum} shows the normalised cumulative distribution of RC compared with that of the RRab population. 
The distribution of RC shows a clear convergence at $R \sim 10^\circ$, while the RRab keeps rising outwards. 
The half-population radius for the RC stars is $R_{H \: RC}=4.2^\circ$ ($R_{H \: RC}=0.61$ kpc) that is nearly half of the value obtained for the RRab distribution ($R_{H \: RRab}=6.8^\circ$).
Therefore, we suggest that the RC stars are more concentrated than the RRab in the mean, but in the central $R<2.2^\circ$ the RRab population is more peaked.  

\begin{figure}
\centering
\includegraphics[scale=0.55]{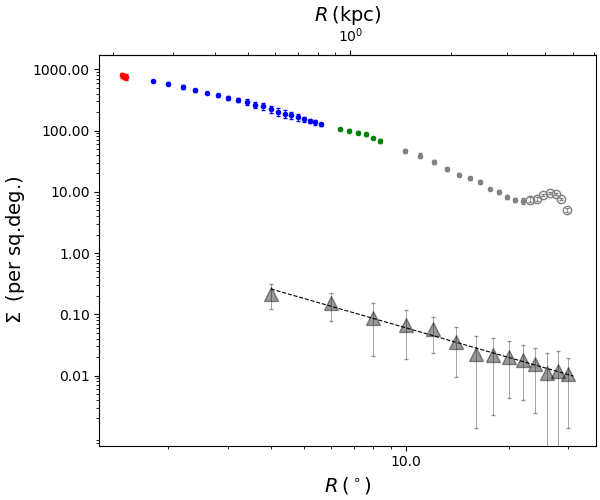}
\caption{
Projected density distribution for RRab (circles) and GCs (triangles) from \cite{harris10} and \cite{Baumgardt}.
The colours of the RRab distribution are the same as in Fig.~\ref{map}. 
The black dashed line shows the best fit with a slope $m_{GCs} =-1.59 \pm 0.060$ for $4.0^\circ<R<30.0^\circ$. 
 }
\label{gc}%
\end{figure}

  Figure~\ref{gc} shows the GC density profile compared to the RRab profile. 
The GC sample follows a power law $\Sigma_{GCs}  \propto R^{-1.59 \pm 0.060}$ for $4.0^\circ<R<30.0^\circ$, which is similar to the RRab projected density distribution in the region $2.2^\circ<R<8.0^\circ$ ($\Sigma_{RRab[2]}  \propto R^{-1.50\pm 0.019}$) and shallower than the slope of the RRab distribution within $8.0^\circ<R<30.0^\circ$ ($\Sigma_{RRab[3]}  \propto R^{-2.43\pm 0.043}$).
It was earlier recognised that the projected density distribution of inner halo RRL stars is similar to that of the GCs  (e.g. \citealt{f80},  \citealt{f82}, and \citealt{d95}).

We separate the GC sample into two groups according to their metallicity, considering that the metallicity distribution of GCs in the bulge is clearly bimodal (\citealt{d95},  \citealt{Barbuy98},  \citealt{Cote00},  \citealt{Fan08}, and \citealt{Forbes10}). 
We find that the density profile for GCs with $[\rm{Fe}/\rm{H}]>-1.1$ dex is $\Sigma_{GCs>}  \propto R^{-1.98 \pm 0.117}$, while for the more metal-poor ($[\rm{Fe}/\rm{H}]<-1.1$ dex) GCs it is $\Sigma_{GCs<}  \propto R^{-1.32 \pm 0.079}$.

The more metal-poor GCs follow a shallower distribution than that obtained for the complete sample similar to the distribution for the RRab population in the same region ($\Sigma_{RRab[2]}  \propto R^{-1.50\pm 0.019}$ and $\Sigma_{RRab[3]}  \propto R^{-2.43\pm 0.043}$), while the GCs with $[\rm{Fe}/\rm{H}]>-1.1$ dex follow a steeper law more consistent with the distribution of the bulge RC ($\Sigma_{RC[2]}  \propto R^{-1.66\pm 0.100}$ and $\Sigma_{RC[3]}  \propto R^{-3.41 \pm 0.075}$).

\begin{figure}
\centering
\includegraphics[scale=0.55]{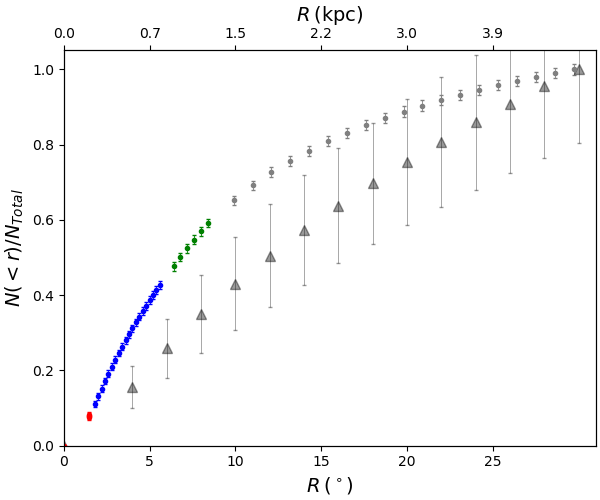}
\caption{
Cumulative distribution for GCs from \cite{harris10} and \cite{Baumgardt} (triangles) and RRab (circles).
The half-population radius for the GCs population is $R_{H \: GCs}=11.9^\circ$ ($R_{H \: GCs}=1.75$ kpc). 
}
\label{gc_cum}%
\end{figure}

The cumulative distribution for GCs compared with the RRab is shown in Fig.~\ref{gc_cum}.
Both distributions follow a similar behaviour. 
The half-population radius for the GCs population is $R_{H \: GCs}=11.9^\circ$ ($R_{H \: GCs}=1.75$ kpc). 
This value is significantly larger than the value obtained for the RRab distribution ($R_{H \: RRab}=6.8^\circ$).
This means that the GCs are less centrally concentrated than RRab stars.
We confirm that this behaviour continues to the inner regions of the Galactic bulge, using the compilation of \cite{harris10} and \cite{Baumgardt}. 
These results do not change if we add the tens of new bulge GCs candidates discovered recently (e.g. \citealt{d17}, and \citealt{tali19}).

There are $N=83$ GCs within $R < 30.0^\circ$ in the catalogues of \cite{harris10} and \cite{Baumgardt}. 
This means that there are roughly $N \sim 800$ field RRab stars per GCs in the whole region. 

One immediate conclusion that can be drawn from the comparison between Fig.~\ref{rc_cum} and Fig.~\ref{gc_cum} is that the GCs appear to follow the RRab distribution rather than the more concentrated distribution of RC in the outer region. 
Regretfully, there are very few GCs, and their sample is incomplete in the inner bulge (\citealt{Ivanov05},  \citealt{Ivanov17},  \citealt{Borissova14},  \citealt{Barbuy16}, and \citealt{d17}).
Therefore, RCs become more suitable comparison targets as they are very numerous in the regions explored here.
 
Table~\ref{tab:2} summarises the results for the fitting procedure obtained for the RRab, RC, and GCs along with the radius at half maximum density and the half-population radius.

\section{Bulge--halo transition region}
\label{sec:sec6}
In this context it is interesting to explore how smooth  the bulge halo transition is at $R =2-3$ kpc ($R=14^\circ - 20^\circ$), and the interplay of the bulge with both the $\sim 10^7$ M$_\odot$ NSC \citep{sch14} and the supermassive black hole ($M_{BH} = 4.14 \times 10^6$ M$_\odot$; \citealt{grav18}) at the centre of our Galaxy. 
To be more specific, we aim to investigate whether a jump in the number of old stars or a change in slope of the density law is present in the transition between these regions.

\begin{figure}
\centering
\includegraphics[scale=0.55]{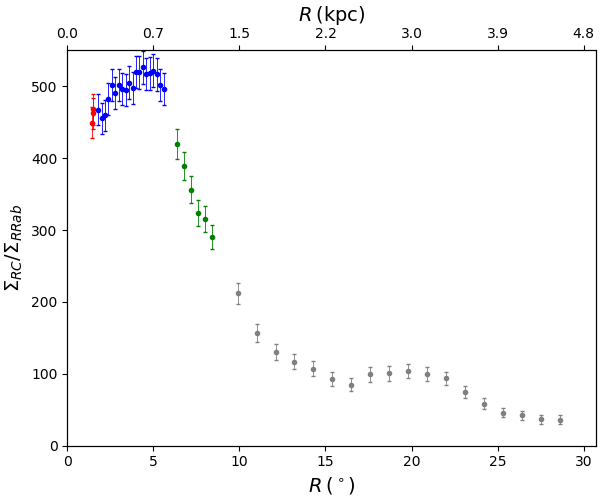}
\caption{
Ratio of the density profile for the RC population and RRab: $\Sigma_{RC}/\Sigma_{RRab}$.
}
\label{rc_comp}%
\end{figure}

We find from Fig.~\ref{ddlog} that the bulge--halo transition region between 2 and 3 kpc is smooth, without a jump in the RRab number density or a significant change in slope of the density power law. 
A more complex behaviour than that suggested by the smooth density distributions of RRab and RC is detected by inspecting the ratio between the number densities of RC and RRab as function of the projected galactocentric distance $R$ (Fig.~\ref{rc_comp}).

We find a central concentration of RRL inside $R \leq 4.0^\circ$ ($R \leq 0.58$ kpc) where the rate constantly decreases as we approach to the centre meaning that the RRL density distribution grows faster than the RC in the innermost region.
Then there is a narrow plateau for the main bulge out to $R \sim 5.2^\circ$ ($R \sim 0.75$ kpc)
after which the ratio decreases until $R \sim 16.5^\circ$ (R $\sim 2.46$ kpc) where it becomes constant again.
The bulge RC giants dominate in the region with $R < 5.2^\circ$ ($R < 0.75$ kpc), where the $N_{RC}/N_{RRab}$ value is five times higher than the outer value.
The distributions of RC and RRab appear to have a similar behaviour in the halo, outside of $R \sim 24^\circ$ where the asymptotic $\Sigma_{RC}/\Sigma_{RRab}$ value is more or less constant, with $\sim 80$ RC stars per RRab. 
The RC distribution diminishes in the outer regions, as does the RRab distribution.
From Fig.~\ref{rc_comp} we claim that the bulge--halo transition covers the range from $\sim1$ kpc until $\sim 2.4$ kpc.
This transition corresponds approximately to the change in slope of the projected density distribution shown in Fig.~\ref{dd} and Fig.~\ref{ddlog} at $R = 1.17$ kpc ($R = 8.0^\circ$). 
The main factor that causes the $\Sigma_{RC}/\Sigma_{RRab}$ to decrease steeply as we approach the Galactic plane and centre, i.e. inside $R \sim 5.2^\circ$ ($R \sim 0.75$ kpc), is that although the RRab and RC distributions change slopes at about $R = 2.2^\circ$ becoming flatter in the inner regions, the RRab distribution shows a steeper slope than that for RC stars. 
Additionally, there are external effects that might contribute to the ratio variance, such as the  presence of disc RC stars that rapidly increases at these low latitudes and the fact that while the RC density profile is well determined in this inner region, we caution that the RRab number counts are quite uncertain as we approach the Galactic centre where optical surveys are more incomplete. Therefore, a deeper survey with the Nancy Roman Space Telescope is needed, as advocated in the next section.

It appears that the presence of the Sgr dwarf galaxy does not affect  the results much. 
On the one hand, this is part of the halo and it is visible as a wide peak at $R \sim 23^\circ$ of the RRab projected density distribution shown in Fig.~\ref{ddlog}, Fig.~\ref{rc}, and Fig.~\ref{gc}.
On the other hand, it is easy to take into account this contribution just by doing a magnitude cut as the Sgr RRL are far behind the bulge (at $25$ kpc compared with $8$ kpc on the mean).

The shape of the ratio $\Sigma_{RC}/\Sigma_{RRab}$  applying the flattening factor remains similar, but the distribution shifts further out, indicating that the results are quite robust against such a flattened model. 

\section{Nancy Roman Space Telescope multi-epoch campaign to search for RR Lyrae stars across the Galactic centre}
\label{sec:sec7}
The most uncertain and at the same time relevant region to investigate the RRL projected density distribution is the dense environment around the Galactic centre, a region that is poorly constrained by current observations, and limited in depth and resolution due to reddening and crowding.

The Nancy Roman Space Telescope, formerly known as the Wide-Field InfraRed Survey Telescope or WFIRST (\citealt{Green12}, and \citealt{Spergel15}), would have the capability to solve this limitation.
The 18 detectors of the Nancy Roman Wide Field Instrument (WFI), each $4096 \times 4096$ pixels in size, will cover a total field of view of $0.28$ sq.deg. at high resolution.
Although this is not one of the major planned Nancy Roman survey programmes, we argue that, with a relatively inexpensive observing campaign, the Nancy Roman would be able to efficiently map the RRL projected density distribution down to the Galactic centre. 
Even in the most highly obscured regions where extinction can reach $A_{K_s} \sim 2$ mag ($A_V \sim 20$ mag), the faintest RRL stars at $8.3$ kpc are bright enough ($Z \sim 21.5$ mag, $J \sim 20$ mag, $K_s \sim 16$ mag; \citealt{contreras18}) to be measured with relatively short WFI integration times. 
The discovery and characterisation of more than $10^4$ RRL would then be enabled by the Nancy Roman with only nine pointings covering the central $2.5$ sq.deg. central region of the MW. 
This will help us to characterise the population of stars, and thus the density profile, in the innermost region $R_{\Sigma \: RRab} <1.7^\circ$, where according to our results just  $\sim 24 \%$ of the expected RRab population has been detected so far.

The RRL proper motion dispersions in the region are $\sigma_l =3.5$ mas/yr, and $\sigma_l \cos b =3.6$ mas/yr \citep{contreras18}, although we might expect to find kinematically cooler subpopulations from disrupted GCs or dwarf satellite galaxies (\citealt{Carlberg17},  \citealt{Barbuy18},   \citealt{d18},  \citealt{Khoperskov18},  \citealt{Prez19},  \citealt{Hughes20},  \citealt{marta20}, and \citealt{Forbes20}).
Furthermore, the Nancy Roman astrometry would be so precise 
that, with a couple of epochs separated by a few years, we would be able to reveal the kinematics of these ancient probes.
For example, \cite{Sanderson19} estimate WFI relative astrometry precise to $0.01$ mas at $H_{AB}=21.6$ mag in typical fields for the whole survey microlensing campaign (\citealt{Spergel15},  \citealt{bennett18}, and \citealt{penny19}). 
Since the bulge RRL stars are much brighter and would have multiple epochs of observations, the WFI astrometry would easily allow us to discriminate distinct ancient subpopulations.
Part of the desired dataset could be achieved as a byproduct of the proposed microlensing survey at the Nancy Roman Space Telescope suggested by \cite{Akeson19}.

Therefore, with the new telescopes it would be possible to solve the puzzle of the formation of the bulge, and even the NSC of our Galaxy and of other galaxies. 


\section{Conclusions}
\label{sec:sec8}
Modern large surveys allow us for the first time to map the constituent stellar populations throughout our Galaxy, all the way from the halo to the Galactic centre.
We chose three representative populations: RRab stars and GCs as classical tracers of the old and relatively metal-poor populations, and RC stars as tracers of relatively more metal-rich populations.

The main catalogues used in this work for RRab stars are the near-IR VVV catalogues of  \cite{contreras18}, \citet{dekany18},  \citet{Majaess18} and D. Majaess private comm.; and the optical catalogues of OGLE \citep{sos14} and Gaia DR2 \citep{Clementini19}. 
For the RC we used the data from 2MASS \citep{Cutri03} and VVV Survey \citep{alonso18}, and finally the catalogues from \cite{harris10} and \cite{Baumgardt} for the GCs analysis.

We computed the RRab projected density distribution $\Sigma_{RRab}$ from the Galactic centre to the halo. 
The distribution is well fitted by three power laws over three different radial ranges, from the innermost region ($R<2.2^\circ$) with $\Sigma_{RRab[1]}  \propto R^{-0.94 \pm 0.051}$ to  $\Sigma_{RRab[2]}  \propto R^{-1.50\pm 0.019}$ for $2.2^\circ<R<8.0^\circ$, and one showing a steeper descent ($\Sigma_{RRab[3]}  \propto R^{-2.43\pm 0.043}$)  in the outer region $8.0^\circ<R<30.0^\circ$. 
We measured the RRab radius at half maximum density and obtain $R_{\Sigma \: RRab} <1.7^\circ$ ($R_{\Sigma \: RRab} <0.25$ kpc). 
Owing to the fact that the distribution follows a power law, this value is an upper limit.
The RRab half-population radius is $R_{H \: RRab}  = 6.8^\circ$ ($R_{H \: RRab}  = 0.99$ kpc). 
The estimated total number of fundamental mode RRL stars in the innermost area within $R<35$ pc is $N \sim 1,562$.
This means that we have discovered less than $1 \%$ of the total number of expected RRab stars in this region.

From the comparison with bulge RC stars we find that the RC projected density distribution has a radius at half maximum density $R_{\Sigma \: RC} <1.6^\circ$ ($R_{\Sigma \: RC} <0.24$ kpc) and half-population radius of $R_{H \: RC} =4.2^\circ$ ($R_{H \: RC} =0.61$ kpc).
The radius at half maximum density for the RC distribution is slightly smaller than the value obtained for the RRab distribution ($R_{\Sigma \: RRab} <1.7^\circ$); however, from the density profile we find that the RC follow a shallower power law ($\Sigma_{RC[1]}  \propto R^{-0.64\pm 0.133}$) than the RRab ($\Sigma_{RRab[1]}  \propto R^{-0.94 \pm 0.051}$) in the most central region ($R<2.2^\circ$). 
Therefore, the RC stars are more concentrated than the RRab throughout the bulge, but in the central $R<2.2^\circ$ the RRab population is more peaked.  

The GC projected density distribution and cumulative distribution show that the GCs are less centrally concentrated than the RRab stars, as expected if a fraction of them have been dynamically disrupted. 
When dividing the sample of GCs into two groups of different metallicities, we find that the more metal-poor ($[\rm{Fe}/\rm{H}]<-1.1$ dex) GCs follow a shallower profile ($\Sigma_{GCs<}  \propto R^{-1.32 \pm 0.079}$) to that of the RRab in the same region.
These results agree with previous studies (\citealt{f82}, and \citealt{d95}). 

Apart from the fact that RRab are ubiquitous members of the MW, present everywhere from the halo to the Galactic centre, we find that they are very concentrated in the innermost region ($R<2.2^\circ$), even more so than the RC population. 
This is an imprint of the MW formation, as RRab stars trace the oldest known stellar populations.

In the innermost region of the Galactic NSC ($R<4.2$ pc) we estimate the presence of $N \sim 168$ stars.
This number is substantially greater than the number proposed by \cite{dong17}; 
 therefore, we claim that the infall and merger of GCs  should contribute to the NSC formation for much more than the $18\%$ of the mass as proposed by \cite{dong17}. 
It should be noted that the dense Galactic central region is still poorly constrained in their RRL content with the observations presently available, and that the Nancy Roman Space Telescope   would be the ideal tool for  completing the RRL census in this crowded region.

In this paper we used the present observational data to adequately describe the behaviour of the RRab, RC, and GC populations as a function of projected galactocentric distance. 
Simply counting these different tracers allowed us to draw some interesting conclusions. 
In a forthcoming paper we will evaluate the results presented in this study in comparison with some models for the formation of the central region of our Galaxy, with special reference to the well-established  {infall and merger} model and the {in situ} scenario. 
The 3D spatial distribution is also proposed for future studies. This is not an easy task because it requires a more complete knowledge of the 3D extinction maps, which will be completed by the new generation of telescopes, and a more specific analysis for the deprojection procedure from a theoretical point of view.

Our Galaxy is currently the only example of a normal large spiral galaxy where we have mapped a wide concentration of old and metal-poor RRab stars up to its centre. 
These results may  provide evidence of the formation process of spiral galaxies in the Universe. 
Given that the Local Group of spirals M31 and M33 are inaccessible, other  galaxies where a proper RRL census can be traced is within lower luminosity galaxies of the Local Group, such as the Sgr dwarf galaxy and the LMC. Nonetheless, their RRL distributions do not appear to be as concentrated as that in the MW.

\section*{Acknowledgements}
We gratefully acknowledge the use of data from the ESO Public Survey program IDs 179.B-2002 and 198.B-2004 taken with the VISTA telescope and data products from the Cambridge Astronomical Survey Unit. 
Support for the authors is provided by the BASAL Center for Astrophysics and Associated Technologies (CATA) through grant AFB 170002, by the Programa Iniciativa Cientifica Milenio grant IC120009, awarded to the Millennium Institute of Astrophysics (MAS), and by Project FONDECYT No. 1170121 and Project FONDECYT No. 1201490.


\begin{table*}
\caption{Catalogues used for this work with the number of RRab stars and radial coverage in  projected galactocentric distance for each catalogue in degrees and parsecs.}  
\label{tab:1}    
\centering                  
\begin{tabular}{c c  c c}      
\hline\hline          
Survey & Number of stars  & Radius coverage & Radius coverage  \\
  &  & \tiny{($^\circ$)} & \tiny{(kpc)} \\
\hline  
VVV1 &  960 & 1.7 &   0.24 \\
VVV2 &  14,882  &  10  &  1.46 \\
OGLE & 27,480   &  8  &  1.17 \\
Gaia   &  21,528 & 30  & 4.80 \\ \hline
Total   &  64,850 & 30 &  4.80 \\
\hline 
\end{tabular}
\end{table*}

\begin{table*}
\caption{Column 1: object type; Col. 2: fitting formulas for the density profiles with (Col. 3) their   radius ranges; Col. 4: radius at half maximum density ($R_\Sigma$); and (Col. 5) half-population radius ($R_H$).} 
\label{tab:2} 
\centering     
\begin{tabular}{c c c c c} 
\hline\hline   
Type & Fitting Law ($\Sigma \propto$)  & Radius range  & Radius at half maximum density ($R_\Sigma$)  & Half-population radius ($R_H$) \\
 &   &   &\tiny{($^\circ$)}  &\tiny{($^\circ$)} \\
\hline   
RRab & $R^{-0.94 \pm 0.051 }$   &   $0^\circ<R<2.2^\circ$ & >1.7 & >6.8  \\
& $R^{-1.50 \pm 0.019}$   &   $2.2^\circ<R<8.0^\circ$ & &   \\
& $R^{-2.43  \pm 0.043}$   &   $8.0^\circ<R<30.0^\circ$ & &   \\

RCs & $R^{-0.64 \pm 0.133 }$   &   $0^\circ<R<2.2^\circ$ & >1.6 & >4.2  \\
& $R^{-1.66 \pm 0.100 }$   &   $2.2^\circ<R<6.5^\circ$ & &   \\
& $R^{-3.41  \pm 0.075}$   &   $6.5^\circ<R<30.0^\circ$ & &   \\

GCs &  $R^{-1.59 \pm 0.66}$ &  $0^\circ<R<30.0^\circ$ &   &  >11.9 \\
GCs  $[\rm{Fe}/\rm{H}]>-1.1$ dex &  $R^{-1.98 \pm 0.117}$ &  $0^\circ<R<30.0^\circ$ &   &  \\
GCs  $[\rm{Fe}/\rm{H}]<-1.1$ dex &  $R^{-1.32 \pm 0.079}$ & $0^\circ<R<30.0^\circ$  &  &   \\
\hline    
\end{tabular}
\end{table*}

\end{document}